\documentclass[twocolumn, showpacs, preprintnumbers,amsmath,amssymb,nofootinbib,balancelastpage]{revtex4-2}

\usepackage{color}
\usepackage{bm}
\usepackage{units}
\usepackage{bbm}

\usepackage[caption=false]{subfig}
\usepackage{graphicx}
\usepackage{dsfont}
\usepackage{amsmath}
\usepackage{array}
\usepackage{epstopdf}
\usepackage{enumerate}
\usepackage{youngtab}
\usepackage{tensor}
\usepackage{braket}
\usepackage[normalem]{ulem}
\usepackage{slashed}
\usepackage[aligntableaux=center]{ytableau}
\usepackage[utf8]{inputenc}
\usepackage[
      colorlinks=true,
      linkcolor=blue,
     urlcolor=blue,
      filecolor=black,
      citecolor=red,
      ]{hyperref}
\usepackage{cleveref}
\usepackage{braket}
\usepackage{mathtools}
\usepackage{rotating}
\usepackage{tabularx}
\newcolumntype{Y}{>{\centering\arraybackslash}X}
\newcolumntype{C}[1]{>{\centering\arraybackslash}p{#1}}

\usepackage{todonotes}
\usepackage{xcolor}
\definecolor{LightCyan}{rgb}{0.7,1,1}
\definecolor{Gray}{gray}{0.9}
\usepackage{xspace}

\begin{document}

\title{Precise Low-Temperature Expansions for the Sachdev-Ye-Kitaev model}

\author{Erick Arguello Cruz and Grigory  Tarnopolsky}
\affiliation{Department of Physics, Carnegie Mellon University, Pittsburgh, PA 15213, USA}


\begin{abstract}
We solve numerically the large $N$ Dyson-Schwinger equations for the Sachdev-Ye-Kitaev (SYK) model
utilizing the Legendre polynomial decomposition and reaching $10^{-36}$ accuracy. 
Using this  we compute the energy of the SYK model at low temperatures $T\ll J$ and obtain its
series expansion up to $T^{7.54}$. While it was suggested that the expansion contains terms $T^{3.77}$ and $T^{5.68}$, we find that  the first non-integer power of temperature 
is $T^{6.54}$, which comes from the two point function of the fermion bilinear operator $O_{h_{1}}=\chi \partial_{\tau}^{3}\chi$
with scaling dimension $h_{1}\approx 3.77$. The coefficient in front of $T^{6.54}$ term agrees well
with the prediction of the conformal perturbation theory. We conclude that the conformal perturbation theory 
appears to work even though the SYK model is not strictly conformal.

\begin{flushright}
\textit{In memory of Yaroslav Pugai}
\end{flushright}

\end{abstract}


\maketitle
\nopagebreak

\section{Introduction}
The Sachdev-Ye-Kitaev (SYK) model is a quantum system of $N$ Majorana fermions with all-to-all random interactions  
 \cite{Kitaev:2015, Sachdev:1992fk}. This model
 finds its application in different branches of science including quantum computation   \cite{Kim:2020luc, PhysRevA.104.012427, Haldar:2020uqn, Hastings:2021ygw}  and there is a variety of proposals for its realization 
in laboratory \cite{PhysRevB.96.121119, PhysRevB.104.035141, PhysRevX.7.031006, PhysRevLett.121.036403, 10.1093/ptep/ptx108, PhysRevA.103.013323, cite-key, PhysRevLett.119.040501, PhysRevA.99.040301, https://doi.org/10.48550/arxiv.2110.04778}.  Reviews of the SYK model can be found in \cite{Maldacena:2016hyu, rosenhaus2019introduction,sarosi2017ads, Chowdhury:2021qpy}. 
 
 The SYK model low temperature behavior displays various interesting properties. Particularly in the large $N$ limit it exhibits  emergent conformal symmetry in the infrared. This suggests that 
 conformal field theory methods can be used to study the model. However 
 the symmetry is both explicitly and spontaneously broken  
 and therefore the term ``nearly conformal quantum mechanics" or $\textrm{NCFT}_{1}$ was coined for the SYK model in \cite{Maldacena:2016hyu}. The conformal symmetry still plays a crucial role and a complete framework of its applicability is not yet clear.  

In this paper we investigate the low temperature expansion of the SYK model 
using the conformal perturbation theory approach and confirm 
theoretical predictions by extremely precise numerical computations. It was previously assumed that the 
SYK model energy at low temperatures contains non-integer powers of temperature $T^{h_{k}}$, 
where $h_{k}$ are scaling dimensions of fermion bilinear operators $O_{h_{k}}$ and the first two dimensions are $h_{1}\approx 3.77$ and 
$h_{2}\approx 5.68$ for $q=4$ case. We show numerically that these terms are not present
 and the first non-integer power of temperature is $T^{2h_{1}-1}$, which comes from the two-point function of the operator  $O_{h_{1}}$. 
 We find that the coefficient in front of this  power term agrees well with the prediction of the conformal perturbation theory. 

The paper is organized as follows:
in section \ref{sectionSYK}, we review the  SYK model. In section \ref{section_Effective_SYK}, we discuss the effective theory  
approach to the SYK model and the low temperature expansion of the two-point function and free energy. In section 
\ref{section_numerical}, we discuss methods for numerical solution of the large $N$  Dyson-Schwinger equations. Finally, in section \ref{section_results}, we present numerical  results and 
compare them with the theoretical predictions.

\section{The SYK model}   
\label{sectionSYK}
The Majorana SYK model is described by the following Hamiltonian  with even integer $q$
\begin{align}
H = \frac{i^{q/2}}{q!}\sum_{i_{1},\dots,i_{q}}J_{i_{1}\dots i_{q}}\chi_{i_{1}}\dots \chi_{i_{q}}\,,
\end{align}
where $N$ Majorana fermions $\chi_{i}$ with $i=1,\dots,N$ interact via 
random couplings $J_{i_{1}\dots i_{q}}$ drawn from a Gaussian ensemble with zero mean 
$\overline{ J_{i_{1}\dots i_{q}}} =0$ and variance $\overline{ J_{i_{1}\dots i_{q}}^{2}} =(q-1)! J^{2}/N^{q-1}$. 
The free energy of the SYK model after disorder average is 
\begin{align}
-\beta F = \overline{  \ln Z }  = \partial_{n}\overline{ Z^{n}}|_{n \to 0}\,,
\end{align}
where $\beta=1/T$ is the inverse temperature.
In \cite{Kitaev:2017awl, Gurau:2017xhf} it was explained that for the SYK model up to $1/N^{q-2}$ order
$\overline{ Z^{n}} = (\overline{Z})^{n}$, therefore at leading $N$  we 
get $-\beta F = \ln \overline{ Z } $.  Taking disorder average of the partition function 
and integrating out the Majorana fermions we obtain a functional integral over two bi-local fields $G$ and $\Sigma$:
\begin{align}
e^{-\beta F}=\int DG(\tau_{1},\tau_{2})D\Sigma(\tau_{1},\tau_{2}) e^{- I[G,\Sigma]}\,,
\end{align}
where $G(\tau_{1},\tau_{2})$ has a meaning of imaginary time two-point function of the Majorana fermions
\begin{align}
G(\tau_{1},\tau_{2})=-\frac{1}{N}\langle \textrm{T} \chi_{i}(\tau_{1})\chi_{i}(\tau_{2})\rangle
\end{align}
and  the $(G,\Sigma)$ action  $I$ is \cite{Maldacena:2016hyu, Kitaev:2017awl}
\begin{align}
&-\frac{ I[G,\Sigma]}{N} =\log \textrm{Pf}(-\sigma-\Sigma) \notag\\
&-\frac{1}{2}\int_{0}^{\beta} d\tau_{1}d\tau_{2} \Big[\Sigma(\tau_{1},\tau_{2})G(\tau_{1},\tau_{2})-\frac{J^{2}}{q}G(\tau_{1},\tau_{2})^{q}\Big] \,, \label{SYKfree}
\end{align}
where $\sigma(\tau_{1},\tau_{2})=\delta'(\tau_{1}-\tau_{2})$.
In the large $N$ limit the leading contribution to the thermodynamic free energy $F$ comes from the saddle point configuration $(G_{*},\Sigma_{*})$ of the functional $I$, so $\beta F_{*}=I[G_{*},\Sigma_{*}]$. This configuration is a solution of the Dyson-Schwinger (DS) equations
\begin{align}
-&\partial_{\tau_{1}}G(\tau_{1},\tau_{2}) - \int_{0}^{\beta} d\tau' \Sigma(\tau_{1},\tau')G(\tau',\tau_{2}) = \delta(\tau_{12})\,, \notag\\
&\Sigma(\tau_{1},\tau_{2}) = J^{2}G(\tau_{1},\tau_{2})^{q-1}\,, \label{SDequations}
\end{align}
and it has translational symmetry $G_{*}(\tau_{1},\tau_{2})=G_{*}(\tau_{12})$. 
It is well-know that the saddle point solution 
for $\beta, \tau_{12} \gg 1/J$ can be approximated by the power law form \cite{Sachdev:1992fk, 2000PhRvL..85..840G, Maldacena:2016hyu, Kitaev:2017awl}
\begin{align}
G_{c}(\tau) = -b^{\Delta}\Big(\frac{\pi}{\beta J |\sin \frac{\pi \tau}{\beta}|}\Big)^{2\Delta}\textrm{sgn}(\tau)\,, 
\label{IRtwopoint}
\end{align}
where $\Delta =1/q$ and 
\begin{align}
b = \frac{1}{2\pi}(1-2\Delta)\tan(\pi \Delta)\,.
\end{align}
This is a consequence of the emergent time reparametrization invariance of 
the DS equations in the IR limit.

Since $\beta F_{*}$ depends only on $\beta J$, the large $N$ energy of the system per particle is $\beta E = J\partial_{J}(\beta F_{*}/N)$ and thus we obtain for the energy 
\begin{align}
E =-\frac{J^{2}}{q}\int_{0}^{\beta}d\tau G_{*}(\tau)^{q}= -\frac{1}{q}\partial_{\tau}G_{*}(\tau)|_{\tau\to 0^{+}} \,.\label{energyE}
\end{align}
This is a UV quantity and cannot be calculated using the conformal approximation (\ref{IRtwopoint}). 
At large $\beta J$, the energy admits a $1/\beta J$ expansion, so 
we have for the  energy per coupling $J$
\begin{align}
\epsilon(\beta J)&=E/J = c_{0}+\frac{c_{2}}{(\beta J)^{2}}+\frac{c_{3}}{(\beta J)^{3}}+\dots\,, \label{Defofeps}
\end{align}
which is a function of $\beta J$ only.
For the first three terms of the energy it was found that \cite{Maldacena:2016hyu, Cotler:2016fpe,  Jevicki:2016bwu, Jevicki:2016ito, Kitaev:2017awl}
\begin{align}
\epsilon(\beta J)= \epsilon_{0}-\frac{\pi^{2}k'(2)\alpha_{0}}{3qa (\beta J)^{2}}+\frac{2\pi^{2}k'(2)\alpha_{0}^{2}}{3a(\beta J)^{3}}+\dots\,, \label{Energyexp}
\end{align}
where $a=((q-1)b)^{-1}$ and  $k(h)$ is given in (\ref{exprforkh}) and  for $q=4$ the ground energy is $\epsilon_{0}\approx -0.0406$ and  $\alpha_{0}\approx 0.2648$, which are determined numerically. The next order terms in the energy expansion are unknown. 
It was suggested in \cite{Cotler:2016fpe,  Kitaev:2017awl} that the next order term should have the form  $c_{h_{1}}/(\beta J)^{h_{1}}$, where $h_{1}$ ($\approx 3.77$ for $q=4$)
is the scaling dimension of the bilinear operator $O_{h_{1}}=\chi_{i} \partial_{\tau}^{3}\chi_{i}$.
On the other hand, the conformal perturbation theory, and  arguments based on the large $q$ expansion in \cite{Tikhanovskaya:2020elb, Tikhanovskaya:2020zcw}, imply that this term should be absent.   The main goal of this paper is to clarify the low temperature energy expansion of the Majorana SYK model. 
For this, we used a high-precision computation of the SYK model Green's function $G_{*}(\tau)$ at very low temperatures (very large $\beta J$).  In the next section we discuss an effective theory approach to the SYK model, which uses conformal perturbation theory for computation of $1/\beta J$ expansion of the two-point function 
and the free energy. 

\section{SYK Effective theory}  
\label{section_Effective_SYK}
The effective theory approach to the SYK model
provides a way to compute higher power temperature corrections (higher $1/\beta J$ terms) of the SYK free energy
and the two point function using the conformal perturbation theory.
In this approach one assumes that the $(G,\Sigma)$ action $I$ can be viewed as a Conformal Field Theory (CFT)  action, perturbed by an infinite series 
of irrelevant bilinear operators $O_{h}(\tau)$
\begin{align}
I =I_{\textrm{CFT}}+\sum_{h}g_{h}\int_{0}^{\beta} d\tau O_{h}(\tau)  \,, \label{CPTaction}
\end{align}
with the scaling exponents $h$ obtained from the solution of the equation $k(h)=1$
with 
\begin{align}
k(h) = \frac{\Gamma (2 \Delta -h) \Gamma (h+2 \Delta -1)}{\Gamma (2 \Delta -2) \Gamma (2 \Delta +1)}\Big(1-\frac{\sin (\pi  h)}{\sin (2 \pi  \Delta )}\Big)\,. \label{exprforkh}
\end{align}
For any $\Delta$ the lowest scaling exponent is $h_{0}=2$, but the higher scaling exponents depend on $q$ and 
for $q=4$ the next three are
\begin{align}
 h_{1}\approx 3.7735, \quad h_{2} \approx 5.6795, \quad h_{3}\approx 7.6320\,. \label{hkformulas}
\end{align}
The bilinear operators $O_{h_{k}}$ schematically are
\begin{align}
O_{h_{k}}(\tau) = \chi_{i}(\tau)\partial_{\tau}^{2k+1}\chi_{i}(\tau)\,. \label{bilocalOper}
\end{align}
Conformal SYK two-point function in (\ref{IRtwopoint}) is obtained using the $I_{\textrm{CFT }}$ action 
\begin{align}
G_{c}(\tau_{12}) = -\frac{1}{Z}\int D\chi_{i}\; \frac{1}{N}\chi_{i}(\tau_{1})\chi_{i}(\tau_{2}) e^{-I_{\textrm{CFT }}}\,.
\end{align}
The $1/\beta J$ corrections to the two-point function
can be computed using conformal perturbation theory with the action in (\ref{CPTaction}).
For the first order correction one uses the three-point function  \cite{Gross:2016kjj, Gross:2017hcz, Gross:2017vhb}
\begin{align}
&\frac{1}{N}\langle \chi_{i}(\tau_{1})\chi_{i}(\tau_{2}) O_{h}(\tau_{3})\rangle =\notag\\
&\;\;=\frac{c_{h}b^{\Delta}\textrm{sgn}(\tau_{12})}{|\frac{\beta J}{\pi}\sin \frac{\pi\tau_{12}}{\beta }|^{2\Delta-h}|\frac{\beta J}{\pi}\sin \frac{\pi\tau_{13}}{\beta }|^{h}|\frac{\beta J}{\pi}\sin \frac{\pi\tau_{23}}{\beta }|^{h}}\,,
\end{align} 
where the structure constants $c_{h}$ can be extracted from the explicit computation of the 
connected part of the four-point function of the Majorana fermions  \cite{Maldacena:2016hyu,Gross:2016kjj}
\begin{align}
c_{h}^{2}= \frac{a}{k'(h)} \frac{(h-1/2)}{ \pi \tan(\pi h/2)}\frac{\Gamma(h)^{2}}{\Gamma(2h)}\,.
\label{chstrconst}
\end{align} 
The second order of the conformal perturbation theory requires the computation of the four-point function and so on \cite{Tikhanovskaya:2020elb}, this in turn uses the structure constants $c_{hh'h''}$ of the three-point function of the operators $O_{h}$, $O_{h'}$ and $O_{h''}$.
These structure constants were computed in \cite{Gross:2017hcz} from the six-point correlation function of the Majorana fermions.
Finally this  leads to the following expansion of the exact two-point function
\begin{align}
G_{*}(\tau) =& G_{c}(\tau) \bigg(1-\sum_{h}\frac{\alpha_{h}}{(\beta J)^{h-1}}f_{h}(\tau)\notag\\
&-\sum_{h,h'}\frac{a_{hh'}\alpha_{h}\alpha_{h'}}{(\beta J)^{h+h'-2}}f_{h,h'}(\tau)-\dots\bigg)\,, \label{Gres}
\end{align}
where the functions $f_{h}(\tau)$ are 
\begin{align}
f_{h}(\tau) = \frac{(2\pi)^{h-1}\Gamma(h)^{2}(A_{h}(e^{\frac{2\pi i \tau}{\beta}})+A_{h}(e^{-\frac{2\pi i \tau}{\beta}}))}{2\sin\frac{\pi h}{2}\Gamma(2h-1)}\,,\label{fhexp}
\end{align}
with $A_{h}(z)=(1-z)^{h}\textbf{F}(h,h,1;z)$ and $\textbf{F}$ is the regularized hypergeometric function.
Parameters $\alpha_{h}$ are related to the couplings $g_{h}$ as 
\begin{align}
g_{h}^{2} = \frac{J^{2} k'(h)}{a} \frac{(h-1/2)}{\pi \tan(\pi h/2)}\frac{\Gamma(h)^{2}}{\Gamma(2h)}\alpha_{h}^{2}
\label{alphahgh}
\end{align}
and the coefficients $a_{hh'}$, $a_{hh'h''}$, etc can be computed exactly  \cite{Tikhanovskaya:2020elb}.  
The functions $f_{h,h'}(\tau)$ are  unknown and depend on $q$,  but scale as $f_{h,h'}(\tau) \to (\beta/\tau)^{h+h'-2}$ for $\beta \to \infty$.
For the Majorana SYK model the leading terms in this series expansion are
\begin{align}
G_{*}(\tau)& =G_{c}(\tau) \bigg(1-\frac{\alpha_{0}}{\beta J}f_{0}(\tau)-\frac{a_{00}\alpha_{0}^{2}}{(\beta J)^{2}}f_{00}(\tau)-\notag\\
&-\frac{\alpha_{1}}{(\beta J)^{h_{1}-1}}f_{1}(\tau)-\frac{a_{000}\alpha_{0}^{3}}{(\beta J)^{3}}f_{000}(\tau)-\dots\bigg)\,. \label{Gres2}
\end{align}
In this approach, the coupling constants $g_{h}$ (or parameters $\alpha_{h}$) are unknown real numbers.
We can find  $\alpha_{h}$ by fitting numerical solution for $G_{*}(\tau)$ by the formula \eqref{Gres2}.

Similarly, using the conformal perturbation theory for the  action \eqref{CPTaction}, we can obtain the $1/\beta J$ expansion of the SYK model free energy 
\begin{align}
    \beta F_{*} &= \beta F_{\textrm{CFT}} + \sum_{h} g_{h}\int_{0}^{\beta} d\tau \langle O_{h}\rangle_{\beta} \notag\\
    &-\frac{1}{2}\sum_{h}g_{h}^{2} \int_{0}^{\beta}d\tau_{1}d\tau_{2}\langle O_{h}(\tau_{1})O_{h}(\tau_{2})\rangle_{\beta}+\dots\,,
\end{align}
and $\beta F_{\textrm{CFT}}/N = \beta E_{0} - s_{0} $ is the conformal part of the free energy, where $E_{0}$ is the bare energy  and $s_{0}$ is the
zero-temperature entropy \cite{2000PhRvL..85..840G,Gu:2019jub}
\begin{align}
s_{0} = \int_{0}^{1/2-\Delta} dx \frac{\pi x}{\tan(\pi x)}\,. \label{zeroTentr}
\end{align}
Since in one-dimension we can map a line to a circle by the transformation $\tau \to e^{i 2\pi \tau/\beta}$ we expect 
that correlation functions on the thermal circle are determined by the conformal symmetry. 
Therefore all one-point correlation functions of primary operators should vanish, except for the identity \cite{Iliesiu:2018fao}. Using the two-point function of operators 
\begin{align}
\langle O_{h}(\tau_{1})O_{h}(\tau_{2})\rangle_{\beta} =N \bigg(\frac{\pi}{\beta J \sin \frac{\pi \tau_{12}}{\beta}}\bigg)^{2h}
\end{align}
we can compute its contribution to the free energy \cite{Tikhanovskaya:2020elb, Maldacena:2016upp} 
\begin{align}
\frac{\beta \delta^{2}F_{h}}{N} = -\frac{\pi^{2h-\frac{1}{2}}\Gamma(\frac{1}{2}-h)}{2\Gamma(1-h)}\frac{g_{h}^{2}/J^{2}}{(\beta J)^{2h-2}}\,.\label{dF2corr}
\end{align}
For the SYK model this CFT approach has a big caveat. 
The problem is that 
the operator $O_{h_{0}}=\chi_{i} \partial_{\tau} \chi_{i}$ with $h_{0}=2$ is essentially 
the Hamiltonian $H$ of the SYK model and thus  correlation 
 functions with it do not have the CFT  structure. For instance as was shown in
\cite{Maldacena:2016hyu} the two-point function of $O_{h_{0}}$ measures the energy fluctuations and is a constant rather than a power law
\begin{align}
\langle O_{h_{0}}(\tau_{1})O_{h_{0}}(\tau_{2})\rangle_{\beta} = N\frac{c}{\beta^{3}}\,,
\end{align}
where $c$ is the specific heat per particle 
\begin{align}
\frac{c}{2\beta} =-\frac{\pi^{2}k'(2)}{3qa} \frac{\alpha_{0}}{ \beta J}\,, \label{specheat1}
\end{align}
so $\beta F_{*}/N = \beta E_{0} - s_{0}-c/(2\beta)+\dots$. 
One can say that the operator $O_{h_{0}}$ breaks conformal symmetry.  On the other hand 
if we assume that $\langle O_{h_{k}}\rangle_{\beta}=0$ for $k=1,2,3,\dots$, then we can heuristically 
argue that for all $n$
\begin{align}
\langle O_{h_{0}}(\tau_{1})\dots O_{h_{0}}(\tau_{n})O_{h_{k}}(\tau_{n+1})\rangle_{\beta} =0  \label{Oh0withOh1}
\end{align}
because, as we mentioned above $ O_{h_{0}}= H$, and  thus
\begin{align}
&\langle H\dots  H O_{h_{k}}(\tau_{n+1})\rangle_{\beta} =\overline{Z^{-1}(-\partial_{\beta})^{n} \textrm{Tr}(O_{h_{k}}e^{-\beta H}) }\,.
\end{align}
Similarly we should conclude that if $\langle O_{h_{k}} O_{h_{m}}\rangle_{\beta} \propto \delta_{km}$ for $k,m=1,2,3,\dots$, then
\begin{align}
\langle O_{h_{0}} \dots O_{h_{0}} O_{h_{k}} O_{h_{m}} \rangle_{\beta} \propto \delta_{km}\,. \label{Oh0OhkOhm}
\end{align}
We find below that our numerical results indeed support the  equations (\ref{Oh0withOh1}) and (\ref{dF2corr}) 
and show some evidence for (\ref{Oh0OhkOhm}) in the case of one $O_{h_{0}}$ operator and $k=m=1$.

Though the expression (\ref{Gres}) can be obtained using the 
CFT approach and assuming that $O_{h_{0}}$ does not violate the CFT structure of the correlation functions, the coefficients $a_{hh'}$, $a_{hh'h''}$ and the first two terms in (\ref{Energyexp}) of the energy expansion  were computed   using  
more direct method, developed in \cite{Kitaev:2017awl, Maldacena:2016hyu, Jevicki:2016bwu, Jevicki:2016ito}. In the Appendix \ref{appC} we show how to reproduce (\ref{dF2corr}) using this method. 
Nevertheless, this method does not
explain which powers of $1/\beta J$ can be present in  the energy expansion.  Therefore 
in this paper we approach this question using numerical computations. Particularly 
we analyse 
$1/\beta J$ expansion of the energy $\epsilon(\beta J)$ and also Green's function $G_{*}(\beta/2)$ at $\tau=\beta/2$ point.

\section{Numerical solution of the Dyson-Schwinger equations}
\label{section_numerical}
Conventional approach to  numerical solution of the SYK  Dyson-Schwinger equations (\ref{SDequations}) at finite $\beta J$
uses iterations of the equations, starting with the free correlation function and using the Fourier series expansion \cite{Maldacena:2016hyu}
\begin{align}
G(\tau) = \frac{1}{\beta}\sum_{n=-\infty}^{+\infty}G(i\omega_{n})e^{-i\omega_{n}\tau}\,,
\end{align}
where $\omega_{n}=2\pi(n+1/2)/\beta$ are the Matsubara frequencies. In the Matsubara frequency space the first  equation in (\ref{SDequations}) is diagonal 
\begin{align}
(i\omega_{n}-\Sigma(i\omega_{n}))G(i\omega_{n})=1\,.
\end{align}
Numerically, one keeps only  Matsubara modes with  $|n| < N_{M}$ and because at large frequencies the Green's function decays as $G(i\omega_{n})\sim 1/(i\omega_{n})$ the error of the numerical solution scales as $O(N_{M}^{-1})$. To improve accuracy of this approach one uses the knowledge of the first $p$ high-frequency expansion terms of the Green's function $G(i\omega_{n})\approx \sum_{k=1}^{p}g_{k}/(i\omega_{n})^{k}$, where $g_{1}=1$. This  helps to reduce the error to the order  $O(N_{M}^{-p-1})$ \cite{doi:10.1063/5.0003145, 2002BlumerPhd, 2007PhDT180C, PhysRevB.94.195119}. In the Appendix \ref{appB}
we discuss this approach to the Majorana SYK model.

In this paper we use  the Legendre spectral method  \cite{PhysRevB.84.075145, doi:10.1063/5.0003145} to solve numerically the Dyson-Schwinger equations (\ref{SDequations}) much more efficiently. Namely
we  decompose the Green's function and the self-energy in the Legendre polynomials
(there are other orthogonal polynomials which can be used \cite{PhysRevB.98.075127} as well as methods \cite{kaye2021discrete})
\begin{align}
G(\tau) = \sum_{\ell=0}^{\infty}G_{\ell}L_{\ell}(x(\tau))\,, \;\; \Sigma(\tau) =  \sum_{\ell=0}^{\infty}\Sigma_{\ell}L_{\ell}(x(\tau))\,, \label{LegenderExp}
\end{align}
where $x(\tau)=2\tau/\beta-1$. The Legendre coefficients $G_{\ell}$ and $\Sigma_{\ell}$ decay exponentially with $\ell$  \cite{PhysRevB.84.075145} 
and therefore if we retain  only $N_{L}$  Legendre coefficients in the expansion (\ref{LegenderExp}) we expect  numerical solution to have exponentially high accuracy of the order $O(e^{-N_{L}})$.

An inherent particle-hole symmetry of the Majorana fermions leads to the symmetry of  Green's function around $\tau =\beta/2$, so $G(\tau)=G(\beta-\tau)$. This implies that all the odd coefficients of the Legendre decomposition must be zero $G_{2k+1}=0$ for $k=0,1,2,\dots$.  

Using the Legendre polynomial expansion we can find the  energy of the SYK model (\ref{Defofeps}) as
\begin{align}
\epsilon(\beta J)  = \frac{1}{2\beta J}\sum_{k=1}^{\infty}k(2k+1)G_{2k} \,,
\label{energySYK}
\end{align}
and also  the two-point function at  $\tau=\beta/2$ point:
\begin{align}
G(\beta/2) =  \sum_{k=0}^{\infty}\frac{(-1)^{k}}{4^{k}}\frac{(2k)!}{(k!)^{2}}G_{2k}\,,
\label{Gbetaover2}
\end{align}
where we used that $L'_{\ell}(-1)=(-1)^{\ell+1}\ell(\ell+1)/2$  and 
$L_{2k+1}(0)=0$ and 
\begin{align}
L_{2k}(0)=\frac{(-1)^{k}}{4^{k}}\frac{(2k)!}{(k!)^{2}} \,.
\end{align}
Details of the numerical solution with the Legendre polynomial expansion 
are discussed in the Appendix \ref{appA}. In the next section we present our numerical results 
and compare them with the theoretical predictions.

\section{Numerical results}
\label{section_results}
\begin{figure}[h!]
\includegraphics[width=0.46\textwidth]{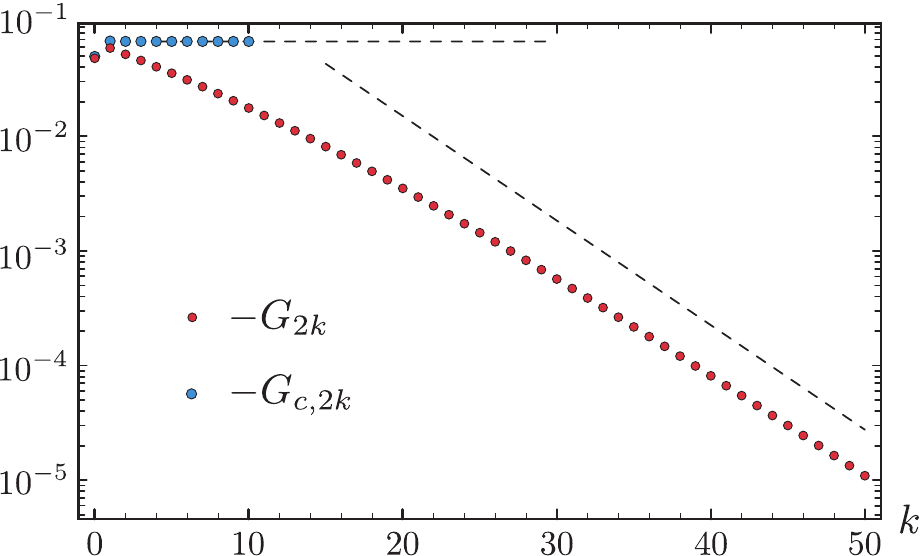}
\caption{Plot of the Legendre coefficients $G_{2k}$ and $G_{c,2k}$ of $G_{*}(\tau)$ and $G_{c}(\tau)$, respectively, for $\beta J =1000$. Two black dashed lines represent asymptotes  (\ref{asymptGn}) and (\ref{Gcasympt}) (Only first $11$ coefficients $G_{c,2k}$ are shown).}
\label{fig3}
\end{figure}

We  find numerical solution for the Green's function $G(\tau)$ in $q=4$ SYK model in the range of parameter $\beta J  \in [10, 15000]$ 
with accuracy around $10^{-36}$ solving  the equations (\ref{SDequations}) with the use of the Legendre polynomials decomposition. To achieve this accuracy we 
retain $N_{L}$ Legendre coefficients, where $N_{L}$ is such that the last Legendre coefficient $G_{N_{L}}$ is smaller than $10^{-36}$. As we already mentioned above the Legendre coefficients $G_{\ell}$ and $\Sigma_{\ell}$ decay exponentially with $\ell$ \cite{PhysRevB.84.075145} and by fitting numerical results for different $\beta J$ we find for $q=4$
\begin{align} 
G_{\ell}\approx -e^{-3.32 \ell/\sqrt{\beta J}}\,, \quad \Sigma_{\ell}\approx -e^{-3.20 \ell/\sqrt{\beta J}} \,. \label{asymptGn}
\end{align}
In contrast, the Legendre coefficients of the conformal solution $G_{c}(\tau)$ in  (\ref{IRtwopoint}) for $q=4$ and large $\ell=2k$  are
\begin{align} 
&G_{c,2k} \approx -\frac{2\sqrt{2}}{\pi^{1/4}\sqrt{\beta J}}\Big(1+\frac{45}{2048 k^{4}}+\dots\Big)\,, \label{Gcasympt}
\end{align}
and to derive it we used expansion in powers of $1-x^{2}$ of the function $(\cos(\pi x/2))^{-2\Delta}$
and the integral 
\begin{align} 
\int_{-1}^{1}\frac{dx L_{2k}(x)}{(1-x^{2})^{\alpha}}= \frac{\Gamma(1-\alpha)\Gamma(k+\frac{1}{2})\Gamma(k+\alpha)}{\Gamma(\alpha)\Gamma(k+1)\Gamma(k+\frac{3}{2}-\alpha)}\,.
\end{align}
A plot of the Legendre coefficients $G_{2k}$ and $G_{c,2k}$ for $\beta J = 1000$ is depicted in Fig.\ref{fig3}. 
We see that only  first two Legendre coefficients $G_{0}$, $G_{2}$ of the exact solution $G_{*}(\tau)$
are close to the ones of the conformal solution $G_{c}(\tau)$.

 We compute the energy using  eq. (\ref{energySYK})  and we can estimate  the accuracy of this computation assuming that each coefficient $G_{\ell}$ for $\ell =1,\dots, N_{L}$ is computed with accuracy $G_{N_{L}}$ plus
 the sum in (\ref{energySYK}) from $N_{L}$ to $\infty$ with the asymptotic values of $G_{\ell}$ from (\ref{asymptGn}).  This gives 
\begin{align} 
\delta \epsilon \approx \Big(\frac{N_{L}^{3}}{24 \beta J}+\frac{N_{L}^{2}}{8\cdot 3.32 \sqrt{\beta J}}\Big)e^{-3.32 N_{L}/\sqrt{\beta J}}\,. \label{Energyerror}
\end{align}
For $\beta J \in [9600, 15000 ] $ we used $N_{L}=3000$
and the lowest accuracy for energy $\epsilon(\beta J)$ is at  $\beta J= 15000$  and can be estimated as $\delta \epsilon \approx 4\cdot 10^{-31}$. On the other hand the coefficient $(1/\beta J)^{7.54} \approx 3 \cdot 10^{-32}$ at this value of $\beta J$, so the overall accuracy is enough to probe $(1/\beta J)^{7.54}$ dependence of the energy. We also 
estimated accuracy of our computation using the second relation in eq. (\ref{Gellconstr}). This estimate shows even higher accuracy of the energy computation than eq. (\ref{Energyerror}), confirming that we can reliably probe $(1/\beta J)^{7.54}$ term.

We are interested in $1/\beta J$ series expansion of the energy $\epsilon(\beta J)$. This is an asymptotic series
and the accuracy of its approximation to $\epsilon(\beta J)$  for large enough $\beta J$ is proportional to $(1/\beta J)^{p'}$,
where $p' > p$ is the next omitted power in the series expansion up to the power $p$. So if we denote the series expansion up to the power $p$ by 
 \begin{align} 
\epsilon_{p}(\beta J) =c_{0}+c_{2}/(\beta J)^{2}+ \dots +c_{p}/(\beta J)^{p}\,,
\end{align}
then we should expect 
\begin{align} 
|\epsilon (\beta J) - \epsilon_{p}(\beta J) | \leq  c_{p'}/(\beta J)^{p'}\,.  \label{asympineq}
\end{align}
This should be true for  asymptotic series provided we take large enough  $\beta J$ such that $c_{p'}/(\beta J)^{p'} \ll c_{p}/(\beta J)^{p}$.  We stress that powers in this expansion can be non-integer numbers. 

\begin{figure}[h!]
\includegraphics[width=0.48\textwidth]{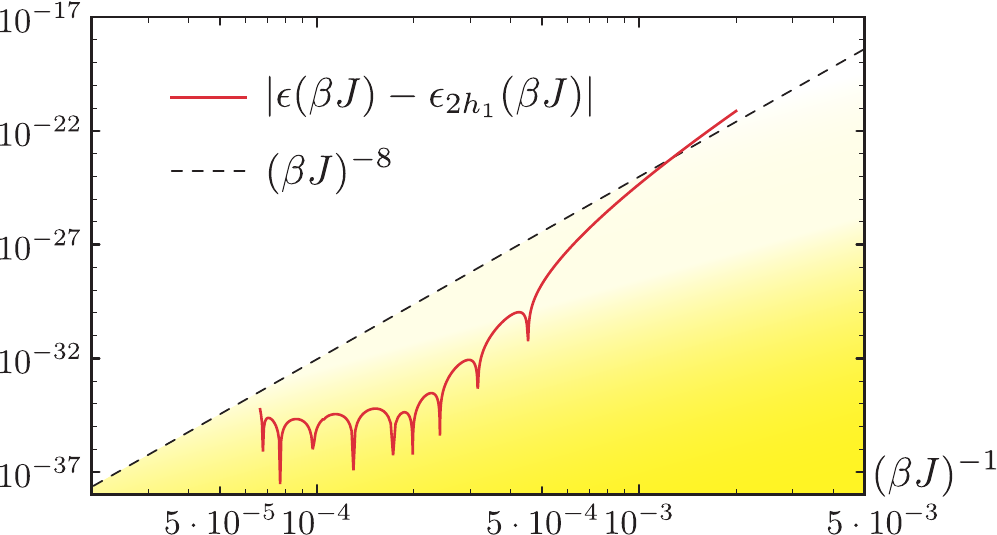}
\caption{Log-log plot of the difference $|\epsilon(\beta J)-\epsilon_{2h_{1}}(\beta J)|$ between the numerical values $\epsilon(\beta J)$ and their fit by the polynomial $\epsilon_{2h_{1}}(\beta J)$, which includes two non-integer powers $2h_{1}-1$ and $2h_{1}$. We tune weights of the loss function (\ref{fittingfun}) such that the difference (red line) lies below $(\beta J)^{-p'}$, where $p'=8$ is the next omitted power in the expansion (black dashed line).  }
\label{fig4}
\end{figure}

In order to extract series coefficients $c_{0}, c_{2}, \dots, c_{p}$ we perform linear regression fit of the numerical data for $\epsilon(\beta J)$ by the polynomial $\epsilon_{p}(\beta J)$ on the interval of couplings $\beta J$ using the weighted least squares, i.e. minimizing the loss function  
\begin{align} 
L_{w}(c_{0},c_{2},\dots, c_{p}) = \sum_{i} w_{i}(\epsilon (\beta J_{i}) - \epsilon_{p}(\beta J_{i}))^{2}  \,.\label{fittingfun}
\end{align}
We tune weights $w_{i}$ in such a way that the inequality (\ref{asympineq}) is fulfilled on the interval of $\beta J$. An example of such a fit on the interval $\beta J \in[500,15000]$ for $p=2h_{1}\approx 7.54$ is depicted in Fig.\ref{fig4}.   

On general grounds, we expect that by fitting the numerical data by the polynomial $\epsilon_{p}(\beta J)$ we can find coefficients $c_{0},c_{2},\dots, c_{p}$ with the accuracy 
\begin{align} 
\delta c_{k} \propto 1/(\beta J_{\textrm{max}})^{p'-k}\,, \label{accuracyck}
\end{align}
where $\beta J_{\textrm{max}}$ is the maximal value of the $\beta J$ interval and $p'>p$ is the next omitted power in the series expansion. The validity of the estimate (\ref{accuracyck}) is not entirely clear, but it 
should be certainly correct in the $\beta J_{\textrm{max}}\to \infty$ limit.
We also analyze accuracy of our fit using the ratio $c_{3}/c_{2}^2$, which is independent on $\alpha_{0}$ parameter and can be computed analytically from (\ref{Energyexp})
\begin{align}
    \frac{c_{3}}{c_{2}^2}=-\frac{384}{\pi(2+3\pi)}\approx -10.6988\,. \label{ratioc3}
\end{align}
The accuracy of this ratio is essentially  the accuracy  of the coefficient $c_{3}$,
since $c_{2}$ should have much higher accuracy than $c_{3}$ for large $\beta J_{\textrm{max}}$  according to (\ref{accuracyck}).
To further improve accuracy of the numerical estimate of the coefficients $c_{k}$, we fit the data for different values of $\beta J_{\textrm{max}}$ and then approximate 
the results to $\beta J_{\textrm{max}}\to \infty$ value. 

To rule out potentially possible non-integer power terms $c_{k}/(\beta J)^{k}$
in the  energy expansion, we include 
these terms one at a time in the polynomial $\epsilon_{p}(\beta J)$ 
and by fitting them we find that their coefficients $c_{k}$ are tiny numbers.  We listed the results of such a fit by the polynomial $\epsilon_{2h_{1}}(\beta J)$ in the Table \ref{table1}.
\begin{table}[htbp]
  \centering
  \begin{tabular}{|c|c|}
\hline
    $c_{1}$ & $\approx 10^{-26}$ \\ 
    \hline
    $c_{h_{1}}$ & $\approx 10^{-13}$ \\ 
    \hline
    $c_{h_{1}+1}$ & $ \approx 10^{-9}$ \\ 
    \hline
    $c_{h_{1}+2}$ &$ \approx 10^{-5}$ \\ 
    \hline
    $c_{h_{2}}$ & $\approx 10^{-5}$ \\ 
\hline
  \end{tabular}
  \caption{Fitting results for possible power expansion coefficients. We conclude that these coefficients are absent in the series expansion. We also included non-existing coefficient $c_{1}$ to check  accuracy of the fit. }
  \label{table1}
\end{table}
We conclude that there are no terms with powers $p=h_{1}, h_{1}+1$, $h_{1}+2$ or $h_{2}$,
where $h_{k}$ are given in (\ref{hkformulas}) in the energy $1/\beta J$ expansion. 

\begin{figure}[h!]
\subfloat[]{
    \includegraphics[width=0.223\textwidth]{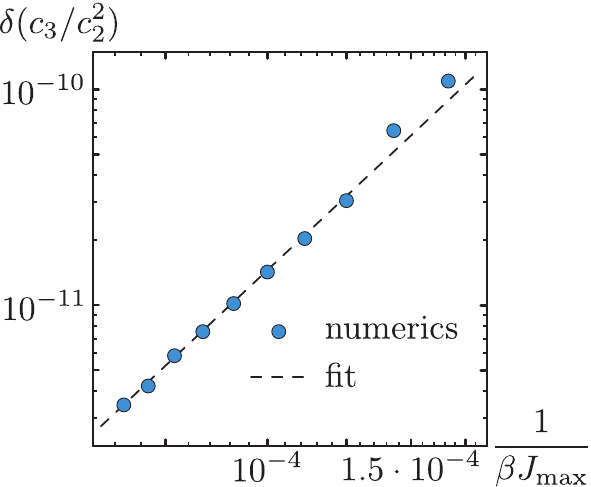}
    \label{fig1a}
}
\subfloat[]{
    \includegraphics[width=0.223\textwidth]{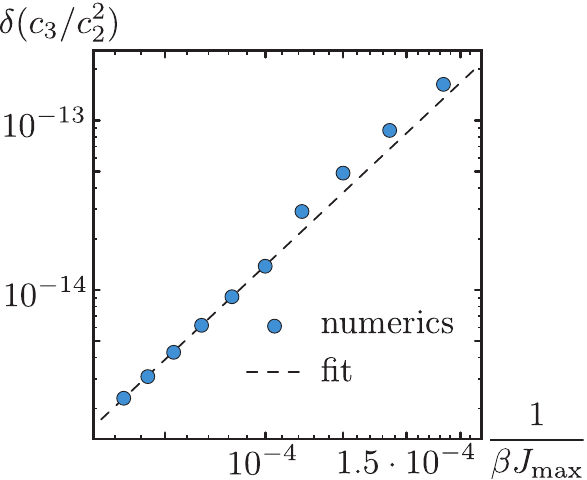}
    \label{fig1b}
}
\caption{Log-log plot of the error $\delta (c_{3}/c_{2}^{2})$ as a function of $1/\beta J_{\textrm{max}}$ obtained from fitting $\epsilon(\beta J)$ by   $\epsilon_{6}(\beta J)$ in (a) and by  
$\epsilon_{7}(\beta J)$ in (b). Black dashed
lines are proportional to $ (\beta J_{\textrm{max}})^{-3.52}$ and $(\beta J_{\textrm{max}})^{-4.44}$.}
\label{Figc3c2}
\end{figure}

To check that non-integer power terms $c_{k}/(\beta J)^{k}$ with $k=2h_{1}-1\approx 6.54$ and $k=2h_{1}\approx 7.54$ are present in the energy expansion we fit the numerical data for $\epsilon(\beta J)$ by the polynomials $\epsilon_{6}(\beta J)$  and 
$\epsilon_{7}(\beta J)$ (with $c_{2h_{1}-1}$ term included in $\epsilon_{7}$) for different values of $\beta J_{\textrm{max}}$ and analyse behavior of the error $\delta(c_{3}/c_{2}^{2})$ as 
a function of $\beta J_{\textrm{max}}$. According to (\ref{accuracyck}) and discussion below (\ref{ratioc3}) we expect to find 
$\delta(c_{3}/c_{2}^{2})\propto 1/(\beta J_{\textrm{max}})^{p'-3}$, where $p'$ is the next omitted power after the power $p$ of the fitting polynomial $\epsilon_{p}(\beta J)$. The plot of  $\delta(c_{3}/c_{2}^{2})$
as a function of $ 1/\beta J_{\textrm{max}}$ and its fit are depicted in Fig. \ref{Figc3c2}. We find numerically $p'\approx 6.52$
and $p'\approx 7.44$ for $p=6$ and $p=7$ cases correspondingly.

Finally we fit the numerical data by the polynomial $\epsilon_{2h_{1}}(\beta J)$ which includes  $c_{2h_{1}-1}$ and $c_{2h_{1}}$ coefficients. 
Our results for the  energy expansion coefficients are summarized in the Table \ref{table2} with the standard error in parenthesis.
\begin{table}[htbp]
  \centering
  \begin{tabular}{|c|c|}
\hline
    $c_{0}$ & $-0.04063026975834491522143475022673(6)$ \\ 
    \hline
    $c_{2}$ & $0.19800839700257657773045(3)$ \\ 
    \hline
    $c_{3}$ & $-0.4194698967373753612(4)$ \\ 
    \hline
    $c_{4}$ & $0.664982720599833(5)$ \\ 
    \hline
    $c_{5}$ & $-2.57685914760(6)$ \\ 
    \hline
    $c_{6}$ & $9.9321852(9)$ \\ 
    \hline
    $c_{2h_{1}-1}$ & $-13.0232(9)$ \\ 
    \hline
    $c_{7}$ & $-26.8(9)$ \\ 
    \hline
    $c_{2h_{1}}$ & $109\pm 5$ \\ 
\hline
  \end{tabular}
  \caption{$1/\beta J$ expansion coefficients of the energy.}
  \label{table2}
\end{table}
For these results the ratio (\ref{ratioc3}) has precision $10^{-18}$.
We will see below  that the coefficient $c_{2h_{1}-1}$ of the first non-integer power term in the energy expansion is in perfect agreement with the prediction of the conformal perturbation theory (\ref{dF2corr}). 

To find numerically the parameters $\alpha_{0}, \alpha_{1}, \alpha_{2}$ we compute $G_{*}(\beta/2)$ on an interval of $\beta J$ using the equation (\ref{Gbetaover2}). We notice that $L_{2k}(0) \approx (-1)^{k}/\sqrt{\pi k} +O(k^{-3/2})$ for $k\gg 1$, so the contribution of the Legendre coefficients $G_{2k}$ in (\ref{Gbetaover2}) decay with $k$, which makes accuracy of $G_{*}(\beta/2)$ computation even higher than the accuracy of the energy computation. 

Similarly to the  energy we fit the numerical results  for $G_{*}(\beta/2)$
by a polynomial of $1/\beta J$, which includes only powers predicted theoretically in (\ref{Gres2}). More precisely to find $\alpha_{0}, \alpha_{1}, \alpha_{2}$ we fit the expression 
\begin{align}
&\frac{\sqrt{\beta J}G_{*}(\beta/2)}{(\pi/4)^{1/4}}+1 =\frac{\alpha_{0}f_{0}(\beta/2)}{\beta J}+\frac{a_{00}\alpha_{0}^{2}f_{00}(\beta/2)}{(\beta J)^{2}}\notag\\
&\qquad+\frac{\alpha_{1}f_{1}(\beta/2)}{(\beta J)^{h_{1}-1}}+\frac{a_{000}\alpha_{0}^{3}f_{000}(\beta/2)}{(\beta J)^{3}}+
\dots\,,
\label{Gbeta/2_expansion}
\end{align}
where we used that $G_{c}(\beta/2)=-(\pi/4)^{1/4}/\sqrt{\beta J}$. Numerical results for the coefficients of the expansion in eq. \eqref{Gbeta/2_expansion} are shown in the Table \ref{table3}. 
\begin{table}[htbp]
  \centering
  \begin{tabular}{|c|c|}
\hline
    \textrm{power of}\; $1/\beta J$  & \textrm{coefficient} \\ 
    \hline
    $1$ & $0.529611247663296071(4)$ \\ 
    \hline
    $2$ & $0.7904019072219(3)$ \\ 
    \hline
   $h_{1}-1$ & $1.548138356(4)$ \\ 
    \hline
    $3$ & $-0.76275663(4)$ \\ 
    \hline
    $h_{1}$ & $9.40092(6)$ \\ 
    \hline
    $4$ & $-11.01(2)$ \\ 
    \hline
    $h_{2}-1$ & $75\pm 20$ \\ 
    \hline
    $h_{1}+1$ & $-140\pm 50$ \\ 
    \hline
    $5$ & $-10\pm 40$ \\ 
    \hline
  \end{tabular}
  \caption{$1/\beta J$ expansion coefficients in eq.  (\ref{Gbeta/2_expansion}).}
  \label{table3}
\end{table}
From these results we can find parameters $\alpha_{h}$  for $h_{0}$, $h_{1}$ and $h_{2}$, using  that $f_{0}(\beta/2)=2$, $f_{1}(\beta/2)\approx 4.7373$, $f_{2}(\beta/2)\approx 11.3971$  from (\ref{fhexp}).
The results for  $\alpha_{h}$ are summarized in Table \ref{table4}.
\begin{table}[htbp]
  \centering
  \begin{tabular}{|c|c|}
\hline
    $\alpha_{0}$ & $0.26480562383164805(2)$ \\ 
    \hline
    $\alpha_{1}$ & $0.32679563(5)$ \\ 
    \hline
    $\alpha_{2}$ & $6.6\pm 1.8$ \\ 
    \hline
  \end{tabular}
  \caption{Parameters $\alpha_{h}$ obtained from  the Table \ref{table3}.}
  \label{table4}
\end{table}
We notice that  numerical error of $\alpha_{2}$ is large, since the two 
powers $h_{2}-1\approx 4.68$ and $h_{1}+1 \approx 4.77$ in $1/\beta J$ expansion are very close to each other and this lowers precision of the fit. 
We also obtain $f_{00}(\beta/2)\approx 5.00969$, $f_{01}(\beta/2)\approx 19.9483$ and $f_{000}(\beta/2) \approx 2.52785$
using that $a_{00}=9/4$, $a_{01}\approx 2.7229$ and $a_{000}=-65/4$ \citep{Tikhanovskaya:2020elb}.

As a check of our accuracy we expect from (\ref{Energyexp})
\begin{align}
c_{2}^{\textrm{theory}}&= \frac{1}{48}\pi (2+3\pi)\alpha_{0}\,.
\end{align}
Using the results from the Tables \ref{table2} and \ref{table4} we find
\begin{align}
c_{2}-c_{2}^{\textrm{theory}}&\approx 10^{-17}\,.
\end{align}
This essentially shows the accuracy of $\alpha_{0}$ from the Table \ref{table4},
since the coefficient $c_{2}$ from the Table \ref{table2}  has higher accuracy in agreement with (\ref{accuracyck}).  Similarly, from (\ref{dF2corr}) we expect 
\begin{align}
c_{2h_{1}-1}^{\textrm{theory}}&=\frac{(h_{1}-1) \pi^{2h_{1}-\frac{1}{2}}\Gamma(\frac{1}{2}-h_{1})}{\Gamma(1-h_{1})}\frac{g_{h_{1}}^{2}}{J^{2}}\notag\\
&\approx -121.9375 \alpha_{1}^{2}\,,
\end{align}
where $g_{h}$ is given in (\ref{alphahgh}).
Using the numerical result for $\alpha_{1}$ from the Table \ref{table4}, we obtain from the conformal perturbation theory  prediction 
\begin{align}
c_{2h_{1}-1}^{\textrm{theory}}& \approx -13.0224\,.
\end{align}
We see that it coincides with our numerical estimate for $c_{2h_{1}-1}$ in the Table \ref{table1} within an error of $8\cdot 10^{-4}$. We remark that such a good agreement with the conformal perturbation theory indirectly 
confirms that  $\langle O_{h_{0}} O_{h_{2}}\rangle = 0$, otherwise it would produce  the power term  $h_{2}+1\approx 6.68$ in the energy expansion, which would strongly interfere with the term $2h_{1}-1\approx 6.54$ 
and thus lower the accuracy of the fit significantly.

\section{Discussion} 
In this article we show numerically that the conformal perturbation theory appears to work for the SYK model even though the SYK model is not strictly
conformal. We found that there are no terms $(\beta J)^{-h_{1}+1}$ and $(\beta J)^{-h_{2}+1}$ in the 
$1/\beta J$ expansion of the SYK free energy and the first non-integer power term  is $p=2h_{1}-2$ ($\approx 5.54$ for $q=4$). It comes from 
the two point function of the bilinear operator $O_{h_{1}} = \chi \partial_{\tau}^{3}\chi$.  
It would be interesting to investigate the low temperature expansion of  the complex SYK model \cite{Sachdev:2015efa}
with non-zero chemical potential $\mu$, where there is a local operator $O_{h}=c^{\dag}\partial^{2}_{\tau}c$ with the scaling dimension 
$h\approx 2.65$ which depends slightly on $\mu$ \cite{Klebanov:2016xxf, Gu:2019jub}. We expect that the first non-integer power in the cSYK model free energy should be $p= 2h-2$ ($\approx 3.3$ for $q=4$) and is much lower than in the Majorana SYK \cite{ArguelloCruzTarnopol}. 
There are variety of other random  and non-random quantum models, where the scaling dimensions of the bilinear operators 
depend strongly on the models' parameters \cite{Kim:2019upg, Klebanov:2020kck, Milekhin:2021sqd, Milekhin:2021cou, Tikhanovskaya:2020elb, Tikhanovskaya:2020zcw}. In some of  these models the scaling dimensions of the lowest operators can be less than $3/2$ (and for some parameters can become even
relevant $h<1$, thus breaking the conformal invariance completely). For these type of models the validity 
of the  conformal perturbation theory approach is of primary importance, since it defines the leading 
behavior of the free energy.

\section*{Acknowledgments}  We would like to thank Igor R. Klebanov and  Alexey Milekhin for useful discussions and comments. 
We also would like to thank Michael Widom
for providing us with computer cluster time.

\appendix

\section{Legendre spectral method}
\label{appA}
In this appendix, following \cite{doi:10.1063/5.0003145}, we briefly describe the Legendre spectral method for solving the Dyson-Schwinger equations.
We would like to solve numerically the Dyson-Schwinger equation 
\begin{align}
\partial_{\tau}G(\tau) +\int_{0}^{\beta}d\tau' \Sigma(\tau-\tau')G(\tau')= -\delta(\tau)\,, \label{DSeq}
\end{align}
where $\Sigma(\tau)=J^{2}G^{q-1}(\tau)$. 
We expand  Green's function $G(\tau)$ in the Legendre series
 \begin{align}
G(\tau) = \sum_{\ell=0}^{\infty}G_{\ell}L_{\ell}(x(\tau))\,,
\end{align}
where $x=2\tau/\beta-1$.
Since the Legendre polynomials $L_{\ell}(x(\tau))$ are defined only on the interval $\tau\in[0,\beta]$($x\in[-1,1]$)
the delta function in (\ref{DSeq}) can be omitted, but its effect must be reinstated in the boundary 
condition for the Green's function\footnote{The delta function implies the boundary condition $G(0^{+})-G(0^{-})=-1$. Using the KMS condition $G(-\tau)=-G(\beta-\tau)$ we arrive to the boundary condition on the interval $\tau \in[0,\beta]$.} $G(0^{+})+G(\beta^{-})=-1$. In terms of the Legendre coefficients this boundary 
condition reads 
\begin{align}
\sum_{\ell=0}^{\infty}((-1)^{\ell}+1)G_{\ell} =-1\,, \label{bcond}
\end{align}
where we used that $L_{\ell}(-x)=(-1)^{\ell}L_{\ell}(x)$ and $L_{\ell}(1)=1$.
For the derivative of the Legendre polynomial we find 
 \begin{align}
\partial_{\tau}L_{\ell}(x(\tau)) = \frac{2}{\beta}\partial_{x}L_{\ell}(x) = \frac{2}{\beta}\sum_{k}D_{k\ell}L_{k}(x)\,,
\end{align}
where $D_{k\ell}=(2k+1)$ if $k+\ell$ is odd and $k<\ell$, otherwise $D_{k\ell}=0$. The derivative matrix $D_{k\ell}$ for the first several values of $k$ and $\ell$ starting from $0$ is 
 \begin{align}
D_{k\ell} = \left(  \begin{array}{cccccc}
    0 & 1 & 0 & 1 & 0 & \hdots \\ 
    0 & 0 & 3 & 0 & 3 & \hdots \\ 
    0 & 0 & 0 & 5 & 0 & \hdots \\ 
    0 & 0 & 0 & 0 & 7 & \hdots \\ 
    0 & 0 & 0 & 0 & 0 & \hdots \\ 
    \vdots & \vdots & \vdots & \vdots & \vdots & \ddots \\ 
  \end{array}\right)\,.
\end{align}
So we obtain that 
\begin{align}
\partial_{\tau}G(\tau)  = \frac{2}{\beta}\sum_{k=0}^{\infty}  \Big[\sum_{\ell=0}^{\infty}D_{k\ell}G_{\ell}\Big] L_{k}(x(\tau))\,.
\end{align}
Next we write the convolution as the Legendre expansion  
\begin{align}
\int_{0}^{\beta}d\tau' \Sigma(\tau-\tau')G(\tau') = \frac{\beta }{2} \sum_{k,\ell=0}^{\infty}[\Sigma*]_{k\ell}G_{\ell}L_{k}(x(\tau))\,.
\end{align}
Therefore the  Dyson-Schwinger equation (\ref{DSeq}) reads
\begin{align}
\sum_{\ell=0}^{\infty}\Big(D_{k\ell}+\frac{\beta^{2}}{4}[\Sigma*]_{k\ell}\Big)G_{\ell} =0 \label{LegDS}
\end{align}
and it has to be supplemented with the boundary condition (\ref{bcond}). We notice that since $G(\tau)=G(\tau/\beta, \beta J)$  the Legendre coefficients $G_{\ell}$ depend only on combination $\beta J$ and this is also evident from the equation (\ref{LegDS}). 

To find  an expression for the convolution matrix $[\Sigma*]_{k\ell}$  we write 
\begin{align}
&\int_{0}^{\beta}d\tau' \Sigma(\tau-\tau')\sum_{\ell=0}^{\infty}G_{\ell}L_{\ell}(x'(\tau')) \notag\\
&\qquad\qquad= \frac{\beta}{2} \sum_{k,\ell=0}^{\infty}[\Sigma*]_{k\ell}G_{\ell}L_{k}(x(\tau))\,,
\end{align}
therefore 
\begin{align}
\int_{-1}^{1}dx' \Sigma(\tau(x)-\tau'(x'))L_{\ell}(x') =   \sum_{k=0}^{\infty}[\Sigma*]_{k\ell}L_{k}(x)\,.
\end{align}
Now multiplying both sides by the Legendre polynomial and integrating over $x$ we obtain 
\begin{align}
[\Sigma*]_{k\ell}&= \frac{2k+1}{2}\int_{-1}^{1}dx'dx  \Sigma(\tau(x)-\tau'(x')) L_{k}(x)L_{\ell}(x') \,,
\end{align}
where we used orthogonality of the Legendre polynomials
\begin{align}
\int_{-1}^{1}dx L_{k}(x)L_{\ell}(x) =\frac{2}{2k+1} \delta_{k\ell}\,.
\end{align}
It is possible to express $[\Sigma*]_{k\ell}$ through the Legendre coefficients $\Sigma_{\ell}$ of $\Sigma(\tau)=\sum_{\ell=0}^{\infty}\Sigma_{\ell}L_{\ell}(\tau(x))$.
The matrix elements of $[\Sigma*]_{k\ell}$ can be found recursively  \cite{doi:10.1137/140955835} 
\begin{align}
[\Sigma*]_{k,\ell+1} =& -\frac{2\ell+1}{2k+3}[\Sigma*]_{k+1,\ell} \notag\\
&+\frac{2\ell+1}{2k-1}[\Sigma*]_{k-1,\ell}+[\Sigma*]_{k,\ell-1}\,. \label{mainrec}
\end{align}
The first two columns of the matrix  $[\Sigma*]_{k,\ell}$ are computed using the relations
\begin{align}
&[\Sigma*]_{k,0} = 
2\Big(\frac{\Sigma_{k-1}}{2k-1}-\frac{\Sigma_{k+1}}{2k+3}\Big), \quad k \geq 1 \notag\\
&[\Sigma*]_{k,1} = \frac{[\Sigma*]_{k-1,0}}{2k-1}-\frac{[\Sigma*]_{k+1,0}}{2k+3}, \quad k \geq 1
\end{align}
with  $[\Sigma*]_{0,0}=-2\Sigma_{1}/3$ and $[\Sigma*]_{0,1}=-[\Sigma*]_{1,0}/3$. The recursion relation (\ref{mainrec}) works only for the lower triangular part 
of the matrix $[\Sigma*]_{k,\ell}$, i.e. for $k \geq \ell$. To compute elements of the upper triangular part one has to use the transpose relation 
\begin{align}
[\Sigma*]_{k,\ell}= (-1)^{\ell+k} \frac{2k+1}{2\ell+1}[\Sigma*]_{\ell,k}\,.
\end{align}
In practice we retain only $N_{L}+1$ Legendre coefficients $G_{\ell}$ and $\Sigma_{\ell}$ with $\ell =0,1,\dots,N_{L}$ and to find $G_{\ell}$ from given $\Sigma_{\ell}$ we 
solve the $(N_{L}+1)\times (N_{L}+1)$ linear matrix equation composed of $k=0,1,\dots, N_{L}-1$ equations (\ref{LegDS}) and $k=N_{L}$ equation (\ref{bcond}).

\begin{figure}[h!]
\includegraphics[width=0.30\textwidth]{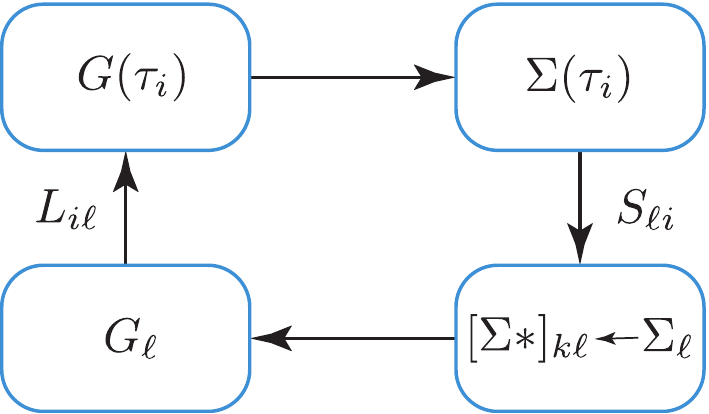}
\caption{Graphic representation of the iteration cycle for the Legendre spectral method. The upper horizontal arrow 
uses  the equation $\Sigma(\tau)=J^{2}G(\tau)^{q-1}$ and the lower horizontal arrow 
uses the equation (\ref{Glupdate}).}
\label{fig1}
\end{figure}

In order to perform fast transform from the Legendre coefficients $G_{\ell}$ to the imaginary time
Green's function $G(\tau)$ we discretize the imaginary time $\tau=\beta(1+x)/2 $ using the  
Legendre-Gauss-Lobatto (LGL) points $x_{i}$ with $i=0,1,2,\dots ,N$, where $x_{i}$ are solutions of the equation
 \begin{align}
(1-x^{2})L'_{N}(x) =0
\end{align}
and they fill the interval $x\in [-1,1]$ with boundary values $x_{0}=-1$ and $x_{N}=1$. The main statement for 
this particular choice of points $x_{i}$ is that the following equation holds for any polynomial $p(x)$ of degree less than $2N$:
 \begin{align}
\int_{-1}^{+1}dx p(x) = \sum_{i=0}^{N}\omega_{i}p(x_{i})\,, \label{LGLstatement}
\end{align}
where  $\omega_{i}=\frac{2}{N(N+1)}\frac{1}{[L_{N}(x_{i})]^{2}}$ are called weights \cite{JieShenSM}.  We assume that we expand $G(\tau)$ in Legendre series up to the $N_{L}$th Legendre polynomial
\begin{align}
G(\tau_{i})=\sum_{\ell=0}^{N_{L}}G_{\ell}L_{\ell}(x_{i}) = \sum_{\ell=0}^{N_{L}}L_{i\ell}G_{\ell}\,,
\end{align}
with $x_{i}=2\tau_{i}/\beta-1$,  therefore the transformation from the Legendre coefficients $G_{\ell}$ to the discretized values of the imaginary 
time Green's function $G(\tau_{i})$ is a multiplication by the matrix $L_{i\ell}=L_{\ell}(x_{i})$. To transform from $G(\tau)$ to $G_{\ell}$ we need to compute the integral
\begin{align}
\int_{0}^{\beta}d\tau G(\tau)L_{\ell}(x(\tau))  = \frac{\beta}{2\ell+1}G_{\ell}\,.
\end{align}
On the other hand if $\ell+N_{L}<2N$ we use (\ref{LGLstatement}) to get
\begin{align}
\int_{0}^{\beta}d\tau G(\tau)L_{\ell}(x(\tau)) =\frac{\beta}{2} \sum_{i=0}^{N} \omega_{i} G(\tau_{i})L_{\ell}(x_{i})\,,
\end{align}
and therefore we find 
\begin{align}
G_{\ell} = \sum_{i=0}^{N} S_{\ell i}G(\tau_{i}), \quad S_{\ell i} = \frac{2\ell+1}{2}\omega_{i}L_{\ell}(x_{i})\,.
\end{align}
In our numerical computations we take $N=N_{L}$. We solve the Dyson-Schwinger equation (\ref{DSeq})
using iterations with the weighted update \cite{Maldacena:2016hyu}:
\begin{align}
&G^{(j+1)}_{\ell} =(1-u)G^{(j)}_{\ell}\notag\\
& +u \cdot \textrm{Solution of}\,[\textrm{eq.} (\ref{LegDS}) + \textrm{eq.} (\ref{bcond}) \; \textrm{with}\; [\Sigma*]_{k\ell}^{(j)} ]\,,
\label{Glupdate}
\end{align}
where $u$ is the weighting parameter and $ [\Sigma*]_{k\ell}^{(j)}$ is constructed from $\Sigma_{\ell}^{(j)}=J^{2}S_{\ell i}G^{(j)}(\tau_{i})^{q-1}$.  We start with the bare solution $G_{\ell}^{(0)}=-\delta_{\ell,0}/2$
and weight $u=1/2$. We monitor the difference $u^{-1}\sum_{i} \Delta\tau_{i} |G^{(j+1)}(\tau_{i})-G^{(j)}(\tau_{i})|^{2}$ between successive iterations 
and divide $u$ by a half if this difference increases.  The iteration cycle is depicted in Fig.\ref{fig1}.
\noindent We performed all computations in Wolfram Mathematica and used small computer cluster
to perform computations for different values of $\beta J$ in parallel.

\section{Matsubara spectral method}
\label{appB}
In this appendix we discuss improved numerical method for solving the Dyson-Schwinger equations (\ref{SDequations}) using the standard Matsubara frequency approach. 
Since in the Majorana SYK model $G(\beta-\tau)=G(\tau)$  Green's function in the 
frequency space is anti-symmetric $G(-i\omega_{n})=-G(i\omega_{n})$ and therefore 
the high frequency expansions of it and of the self-energy have only odd powers of frequency
with real coefficients 

\begin{align}
G(i\omega_{n}) =\sum_{k=1}^{\infty}\frac{g_{2k-1}}{(i\omega_{n})^{2k-1}},\;  
\Sigma(i\omega_{n}) =\sum_{k=1}^{\infty}\frac{s_{2k-1}}{(i\omega_{n})^{2k-1}}
\end{align}
and $g_{1}=1$. Expanding the r.h.s. of the DS equation 
\begin{align}
G(i\omega_{n}) =(i\omega_{n}-\Sigma(i\omega_{n}))^{-1}
\end{align}
for large $i\omega_{n}$ we find relations 
between the coefficients $g_{2k-1}$ and $s_{2k-1}$
\begin{align}
g_{3}= s_{1}, \;\; g_{5} = s_{1}^{2}+s_{3}, \;\; g_{7} = s_{1}^{3}+2s_{1}s_{3}+s_{5},\;\dots \label{gsrel}
\end{align}
Denoting an inverse Fourier transform of a given frequency power as
\begin{align}
Q_{p}(\tau) =\frac{1}{\beta}\sum_{n=-\infty}^{+\infty} \frac{e^{-i\omega_{n}\tau}}{(i\omega_{n})^{p}} =
\mathop{\textrm{res}}_{z=0}\Big(\frac{-e^{z(\beta/2-\tau)}}{2z^{p}\cosh(\frac{\beta z}{2})}\Big),
\end{align}
where the first three functions are $Q_{1}(\tau)=-1/2$, $Q_{2}(\tau)=(2\tau -\beta)/4$, $Q_{3}(\tau)=\tau(\beta-\tau)/4$,
and using the equation $\Sigma(\tau)=J^{2}G(\tau)^{q-1}$ we find  
\begin{align}
s_{1} = J^{2}/2^{q-2}\,,
\end{align}
and thus $g_{3}=J^{2}/2^{q-2}$. This allows to write  expansion of  the Green's function for $J \tau \ll 1$
\begin{align}
G(\tau) = -\frac{1}{2}-q \epsilon J \tau -\frac{(J \tau)^{2}}{2^{q}}+\dots\,,
\end{align}
where we used eqs. (\ref{energyE}) and (\ref{Defofeps}) to write the linear $\tau$ term.  Consequently this fixes a small $J\tau$ expansion of the self-energy
\begin{align}
&\Sigma(\tau) = -J^{2}\bigg(\frac{1}{2^{q-1}}+\frac{q(q-1)}{2^{q-2}}\epsilon J\tau \notag\\
&\;\;+\Big(\frac{q-1}{2^{2q-2}}+\frac{q^{2}(q-1)(q-2)}{2^{q-2}}\epsilon^{2} \Big)(J\tau)^{2}+\dots\bigg)\,.
\end{align}
Since we can write $\Sigma(\tau)$ as 
\begin{align}
\Sigma(\tau)=\sum_{k=1}^{\infty}s_{2k-1}Q_{2k-1}(\tau)\,,
\end{align}
and  only $Q_{3}(\tau)$ contains $\tau^{2}$ term  we find 
\begin{align}
s_{3}= J^{4}\left(\frac{q-1}{2^{2q-4}}+\frac{q^{2}(q-1)(q-2)}{2^{q-4}}\epsilon^{2} \right)
\end{align}
and therefore from (\ref{gsrel}) we obtain 
\begin{align}
g_{5}= J^{4}\left(\frac{q}{2^{2q-4}}+\frac{q^{2}(q-1)(q-2)}{2^{q-4}}\epsilon^{2} \right)\,.
\end{align}
We see that only $g_{1}$, $g_{3}$ and $s_{1}$ coefficients are known explicitly,
whereas already $g_{5}$ and $s_{3}$ contain the energy $\epsilon$, which itself 
can be computed only from the exact solution of the DS equations. Finally we note that the
Matsubara coefficients $G(i\omega_{n})$ are  related to the Legendre coefficients $G_{\ell}$ as \cite{PhysRevB.84.075145}
\begin{align}
G(i\omega_{n})= \sum_{\ell=0}^{\infty}T_{n\ell}G_{\ell}\,, \label{MatFromLeg}
\end{align}
where the matrix $T_{n\ell}$ is 
\begin{align}
T_{n\ell} = \beta (-1)^{n}i^{\ell+1} j_{\ell}(\beta \omega_{n}/2)
\end{align}
and $j_{\ell}(z)$ are the spherical Bessel functions. The matrix $T_{n\ell}$ satisfies 
$\sum_{n}T_{n\ell}^{*}T_{n\ell'} = \beta^{2}\delta_{\ell \ell'}/(2\ell+1)$. 
Using the properties of the spherical Bessel functions one can relate the coefficients $g_{k}$  to the Legendre coefficients $G_{\ell}$ 
\begin{align}
g_{k} = \frac{2(-1)^{k}}{\beta^{k-1}}\sum_{\ell=0}^{\infty}\frac{(\ell+k-1)!}{(k-1)!(\ell-k+1)!}\delta_{\ell+k,\textrm{odd}}G_{\ell}\,. \label{gkFromGl}
\end{align}
This relation imposes constrains on the Legendre coefficients of the Majorana SYK two-point function 
\begin{align}
&\sum_{\ell=0:2}^{\infty}G_{\ell} = -\frac{1}{2}\,, \notag\\
&\sum_{\ell=0:2}^{\infty}(\ell-1)\ell(\ell+1)(\ell+2)G_{\ell} = -\frac{(\beta J)^{2}}{2^{q-2}}\,, \label{Gellconstr}
\end{align}
where we used expressions for $g_{1}$ and $g_{3}$ coefficients. The first equation in (\ref{Gellconstr}) is essentially 
the boundary condition (\ref{bcond}) and our numerical solution for $G_{\ell}$ automatically satisfies it. The second equation in (\ref{Gellconstr}) can be used to estimate  accuracy of the numerical results for $G_{\ell}$ coefficients. 

\begin{figure}[h!]
\includegraphics[width=0.30\textwidth]{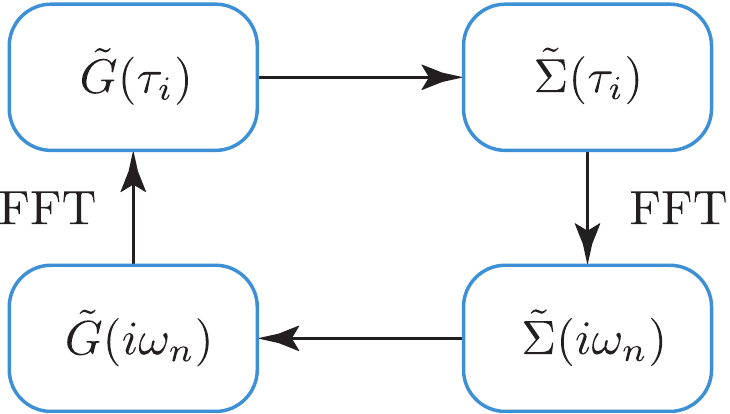}
\caption{Graphic representation of the iteration cycle for the improved Matsubara spectral method.}
\label{fig2}
\end{figure}

To increase numerical accuracy of the Matsubara spectral method we rewrite the DS equations in terms of the 
reduced functions 
\begin{align}
&\tilde{G}(i\omega_{n}) =G(i\omega_{n}) -\frac{1}{i\omega_{n}}- \frac{J^{2}}{2^{q-2}(i\omega_{n})^{3}}\,, \notag\\
&\tilde{\Sigma}(\tau) = \Sigma(\tau)+\frac{J^{2}}{2^{q-1}}\,, \label{GSigReduce}
\end{align}
and $\Sigma(\tau)$ is computed as 
\begin{align}
\Sigma(\tau) = J^{2}\Big(\tilde{G}(\tau)+Q_{1}(\tau)+\frac{J^{2}}{2^{q-2}}Q_{3}(\tau)\Big)^{q-1}\,.
\end{align}
Therefore the fast Fourier transform (FFT) is performed with the functions $\tilde{G}(i\omega_{n})$ and $\tilde{\Sigma}(i\omega_{n})$, which  decay faster than $G(i\omega_{n})$ and $\Sigma(i\omega_{n})$ in the frequency space. 
The iteration cycle is depicted in Fig.\ref{fig2}.

Finally we discuss computation of $\log \textrm{Pf}(\partial_{\tau}-\Sigma)$, which contains 
the zero-temperature entropy (\ref{zeroTentr}) and has  $1/\beta J$ expansion of the form 
\begin{align}
&2\log \textrm{Pf}(\partial_{\tau}-\Sigma) = -(q+1)\beta J \epsilon_{0} +2s_{0} \notag\\
&\qquad\qquad -(q-3)\frac{c_{2}}{\beta J}-(q-2)\frac{c_{3}}{(\beta J)^{2}}-\dots\,,
\end{align}
where $c_{k}$ are the expansion coefficients of the  energy (\ref{Defofeps}). A naive 
computation with the use of the Legendre coefficients 
\begin{align}
2\log \textrm{Pf}(\partial_{\tau}-\Sigma) = \log 2\textrm{det}(1+ \frac{\beta^{2}}{4}D^{-1}[\Sigma *])
\end{align}
has very low accuracy of the order $ O(N_{L}^{-1})$. Notice that in the matrix $D_{kl}$ the last $k=N_{L}$ row is $(-1)^{\ell}+1$ and represents the boundary condition (\ref{bcond}).  To improve accuracy one can compute Matsubara coefficients from the Legendre ones using (\ref{MatFromLeg}) and then compute 
\begin{align}
&\log \textrm{Pf}(\partial_{\tau}-\Sigma) =\frac{1}{2} \log \Big(2\cosh^{2}\Big(\frac{\beta J}{2^{q/2}}\Big)\Big) \notag\\
&\qquad\qquad+ \sum_{n=0}^{N_{M}} \log\Big(1+\frac{i
\omega_{n}\tilde{\Sigma}(i\omega_{n})}{\omega_{n}^{2}+J^{2}/2^{q-2}}\Big)\,,
\end{align}
where $\tilde{\Sigma}(i\omega_{n})$ is the reduced self-energy (\ref{GSigReduce}). We remark that whereas the accuracy of $g_{k}$ computed  from $G_{\ell}$ using (\ref{gkFromGl}) decay
exponentially with $k$ because of the exponentially growing factors, the accuracy of $G(i\omega_{n})$ computed using (\ref{MatFromLeg}) remains exponential.

\section{Computation of the free energy using Kitaev-Suh resonance theory}
\label{appC}

We can rewrite the SYK $(G,\Sigma)$ action (\ref{SYKfree}) in the following form
\begin{align}
&-\frac{I}{N} = \log \textrm{Pf}(-\hat{\Sigma}) -\frac{1}{2}\int_{0}^{\beta} d\tau_{1}d\tau_{2} \Big[\hat{\Sigma}(\tau_{1},\tau_{2})G(\tau_{1},\tau_{2})\notag\\
&-\frac{J^{2}}{q}G(\tau_{1},\tau_{2})^{q}\Big] +\frac{1}{2}\int d\tau_{1}d\tau_{2} \sigma(\tau_{1},\tau_{2})G(\tau_{1},\tau_{2}) \,, \label{Inew}
\end{align}
where we made a replacement $\hat{\Sigma}=\sigma + \Sigma$. The first part of this action without the last term is invariant under reparametrizations of time \cite{Kitaev:2017awl} and can be viewed as $-I_{\textrm{CFT}}/N$ introduced in (\ref{CPTaction}). The saddle point configuration of $I_{\textrm{CFT}}$ is $(G_{c}(\tau), \Sigma_{c}(\tau))$.  
The last term in (\ref{Inew}) is considered as a perturbation.
The leading correction to the free energy is obtained from the first order perturbation theory  
\begin{align}
-\frac{\beta \delta F}{N} & = \frac{1}{2}\int_{0}^{\beta}d\tau_{1}d\tau_{2} \sigma(\tau_{1},\tau_{2})\langle G(\tau_{1},\tau_{2}) \rangle \notag\\
&=\frac{\beta}{2}\int_{0}^{\beta}d\tau \sigma(\tau) G_{c}(\tau) \,.
\end{align}
We can not use direct expression for the source function $\sigma(\tau)=\delta'(\tau)$ in this formula, as it 
leads to uncontrolled divergence. It was proposed in \cite{ Kitaev:2017awl} that the source function $\sigma$ can be replaced by the expression 
\begin{align}
\sigma(\tau) = -\sum_{h}\sigma_{h}\frac{J^{2} \textrm{sgn}(\tau)}{b^{\Delta}a^{1/2}|J \tau|^{h +1-2\Delta}}u(\xi)\,,
\end{align}
where $a=((q-1)b)^{-1}$ and the sum goes over the scaling exponents of the bilinear operators in (\ref{bilocalOper}) and $u(\xi)$ with $\xi = \ln |J\tau| $ is the window function, which serves as a regulator with the normalization $\int d\xi u(\xi) =1$. 
From the computation of the corrections to the two-point function, 
one derives that the coefficients $\alpha_{h}$ in (\ref{Gres}) are related to $\sigma_{h}$ as \cite{Kitaev:2017awl}
\begin{align}
\alpha_{h} = \frac{\sigma_{h}a^{1/2}}{-k'(h)}\,. \label{alphasigma}
\end{align}
Then the contribution to $\delta F$ comes from the 
$h_{0}=2$ scaling exponent and the second order term of $\tau/\beta$ expansion of $G_{c}$:
\begin{align}
&-\frac{\beta \delta F}{N} =-\frac{\beta \sigma_{0} J^{2}}{2b^{\Delta}a^{1/2}}\notag\\
&\quad \times 2\int_{0}^{\beta/2}d\tau  \frac{u(\xi)}{|J \tau|^{3-2\Delta}}\left(-\frac{b^{\Delta}}{|J\tau|^{2\Delta}}\frac{\pi^{2}\Delta}{3} \frac{\tau^{2}}{\beta^{2}}\right)\,, 
\end{align}
where the factor of $2$ in front of the integral is due to the other half of the thermal circle. 
Using  normalization of the window function and (\ref{alphasigma}) 
we find the leading correction to the free energy
\begin{align}
\frac{\beta \delta F}{N} & =\frac{\pi^{2} k'(2)}{3 q a }\frac{\alpha_{0}}{\beta J} \,,
\end{align}
which agrees with (\ref{specheat1}) and (\ref{Energyexp}). 
It is not clear why a similar calculation with the higher  scaling exponents $h$  would not produce $1/(\beta J)^{h-1}$ contribution. 

The next $\delta^{2}F$ contribution to the free energy comes from 
the second order perturbation theory 
\begin{align}
-\frac{\beta \delta^{2} F}{N}  =&\frac{N}{8}\int_{0}^{\beta}d\tau_{1}\dots d\tau_{4} \sigma(\tau_{1},\tau_{2})\sigma(\tau_{1},\tau_{2})\notag\\
&\times \langle  G(\tau_{1},\tau_{2}) G(\tau_{3},\tau_{4})\rangle_{\textrm{conn}} . \label{F2ndorder}
\end{align}
The connected correlation function of two $G(\tau_{1},\tau_{2})$ fields is equal to the connected part of the four point function 
\begin{align}
\langle  G(\tau_{1},\tau_{2}) G(\tau_{3},\tau_{4})\rangle_{\textrm{conn}} =\frac{1}{N}\mathcal{F}(\tau_{1},\tau_{2};\tau_{3},\tau_{4})\,.
\end{align}
The $h_{0}=2$ contribution to the four-point function is special and we only consider $h\neq 2$ sector here. It is well-known that 
in this sector the four-point function can be written as \cite{Maldacena:2016hyu, Gross:2016kjj, Kitaev:2017awl}
\begin{align}
&\mathcal{F}_{h\neq 2}(\tau_{1},\tau_{2};\tau_{3},\tau_{4}) =\notag\\
&=G_{c}(\tau_{12})G_{c}(\tau_{34}) \sum_{k=1}^{\infty}c_{h_{k}}^{2}\chi^{h_{k}} \,_{2}F_{1}(h_{k},h_{k},2h_{k},\chi)\,,
\label{Confblocks}
\end{align}
where $c_{h}$ are given in (\ref{chstrconst})
and the cross-ratio $\chi$ at finite temperature is 
\begin{align}
\chi = \frac{\sin \frac{\pi \tau_{12}}{\beta}\sin \frac{\pi \tau_{34}}{\beta}}{\sin \frac{\pi \tau_{13}}{\beta}\sin \frac{\pi \tau_{24}}{\beta}}\,.
\end{align} 
Taking only $\sigma_{h}$ terms from $\sigma$ functions  and  picking term with the scaling dimension $h$ in (\ref{Confblocks})  
we find contribution to the free energy 
\begin{align}
-&\frac{\beta \delta^{2} F_{h}}{N}  = \frac{J^{4}c_{h}^{2}\sigma_{h}^{2}}{8a}\int_{0}^{\beta}d\tau_{1}\dots d\tau_{4} \notag\\   
&\quad\times \frac{u(\xi_{12})u(\xi_{34})}{|J \tau_{12}|^{h+1}|J \tau_{34}|^{h+1}}\ \chi^{h} \,_{2}F_{1}(h,h,2h,\chi)\,.
\end{align}
Since the window function decays quickly when $\xi = \log |J \tau|$ is not small, we assume that the integrals acquire the main contribution for  $\tau_{12} \to 0$ and $\tau_{34}\to 0$, therefore we use that $\,_{2}F_{1}(h,h,2h,\chi) \approx 1 +\dots$ and get 
\begin{align}
&-\frac{\beta \delta^{2} F_{h}}{N}  = \frac{J^{4}c_{h}^{2}\sigma_{h}^{2}}{8a}4\int_{0}^{\beta/2} dx dx' \notag\\
&\;\;\times \int_{0}^{\beta}dy dy'\frac{u(\xi)u(\xi')}{|J x|^{h+1}|J x' |^{h+1}} \bigg(\frac{\frac{\pi x}{\beta}\frac{\pi x'}{\beta}}{\sin^{2}\frac{\pi (y-y')}{\beta}}\bigg)^{h}\,,
\end{align}
where we introduced $x=\tau_{12}$, $x'=\tau_{34}$ and $y=(\tau_{1}+\tau_{2})/2$, $y'=(\tau_{3}+\tau_{4})/2$. Using normalization of the window function $u(\xi)$, the relation between $\sigma_{h}$ and $\alpha_{h}$ in (\ref{alphasigma}) as well as the one between $\alpha_{h}$ and $g_{h}$ in (\ref{alphahgh}), and the expression for $c_{h}$ in (\ref{chstrconst}), we finally obtain 
\begin{align}
\frac{\beta \delta^{2} F_{h}}{N} & =-\frac{1}{2}g_{h}^{2} \int_{0}^{\beta} dy dy' \bigg(\frac {\pi}{\beta J \sin\frac{\pi (y-y')}{\beta}}\bigg)^{2h}\,.
\end{align}
This is exactly equal to the result of the conformal perturbation theory (\ref{dF2corr}).

\bibliographystyle{ieeetr}
\bibliography{sykthermo}

\begin{thebibliography}{10}

\bibitem{Kitaev:2015}
A.~Kitaev, ``{A simple model of quantum holography},''
\newblock
  \url{http://online.kitp.ucsb.edu/online/entangled15/kitaev/},\url{http://online.kitp.ucsb.edu/online/entangled15/kitaev2/}.
  Talks at KITP, April 7, 2015 and May 27, 2015.

\bibitem{Sachdev:1992fk}
S.~Sachdev and J.~Ye, ``{Gapless spin fluid ground state in a random, quantum
  Heisenberg magnet},'' {\em Phys. Rev. Lett.}, vol.~70, p.~3339, 1993.

\bibitem{Kim:2020luc}
J.~Kim, J.~Kim, and D.~Rosa, ``{Universal effectiveness of high-depth circuits
  in variational eigenproblems},'' {\em Phys. Rev. Res.}, vol.~3, no.~2,
  p.~023203, 2021.

\bibitem{PhysRevA.104.012427}
V.~P. Su, ``Variational preparation of the thermofield double state of the
  sachdev-ye-kitaev model,'' {\em Phys. Rev. A}, vol.~104, p.~012427, Jul 2021.

\bibitem{Haldar:2020uqn}
A.~Haldar, O.~Tavakol, and T.~Scaffidi, ``{Variational wave functions for
  Sachdev-Ye-Kitaev models},'' {\em Phys. Rev. Res.}, vol.~3, no.~2, p.~023020,
  2021.

\bibitem{Hastings:2021ygw}
M.~B. Hastings and R.~O'Donnell, ``{Optimizing Strongly Interacting Fermionic
  Hamiltonians},'' 10 2021.

\bibitem{PhysRevB.96.121119}
A.~Chew, A.~Essin, and J.~Alicea, ``Approximating the sachdev-ye-kitaev model
  with majorana wires,'' {\em Phys. Rev. B}, vol.~96, p.~121119, Sep 2017.

\bibitem{PhysRevB.104.035141}
R.~Haenel, S.~Sahoo, T.~H. Hsieh, and M.~Franz, ``Traversable wormhole in
  coupled sachdev-ye-kitaev models with imbalanced interactions,'' {\em Phys.
  Rev. B}, vol.~104, p.~035141, Jul 2021.

\bibitem{PhysRevX.7.031006}
D.~I. Pikulin and M.~Franz, ``Black hole on a chip: Proposal for a physical
  realization of the sachdev-ye-kitaev model in a solid-state system,'' {\em
  Phys. Rev. X}, vol.~7, p.~031006, Jul 2017.

\bibitem{PhysRevLett.121.036403}
A.~Chen, R.~Ilan, F.~de~Juan, D.~I. Pikulin, and M.~Franz, ``Quantum holography
  in a graphene flake with an irregular boundary,'' {\em Phys. Rev. Lett.},
  vol.~121, p.~036403, Jul 2018.

\bibitem{10.1093/ptep/ptx108}
I.~Danshita, M.~Hanada, and M.~Tezuka, ``{Creating and probing the
  Sachdev–Ye–Kitaev model with ultracold gases: Towards experimental
  studies of quantum gravity},'' {\em Progress of Theoretical and Experimental
  Physics}, vol.~2017, 08 2017.
\newblock 083I01.

\bibitem{PhysRevA.103.013323}
C.~Wei and T.~A. Sedrakyan, ``Optical lattice platform for the
  sachdev-ye-kitaev model,'' {\em Phys. Rev. A}, vol.~103, p.~013323, Jan 2021.

\bibitem{cite-key}
Z.~Luo, Y.-Z. You, J.~Li, C.-M. Jian, D.~Lu, C.~Xu, B.~Zeng, and R.~Laflamme,
  ``Quantum simulation of the non-fermi-liquid state of sachdev-ye-kitaev
  model,'' {\em npj Quantum Information}, vol.~5, no.~1, p.~53, 2019.

\bibitem{PhysRevLett.119.040501}
L.~Garc\'{\i}a-\'Alvarez, I.~L. Egusquiza, L.~Lamata, A.~del Campo, J.~Sonner,
  and E.~Solano, ``Digital quantum simulation of minimal
  $\mathrm{AdS}/\mathrm{CFT}$,'' {\em Phys. Rev. Lett.}, vol.~119, p.~040501,
  Jul 2017.

\bibitem{PhysRevA.99.040301}
R.~Babbush, D.~W. Berry, and H.~Neven, ``Quantum simulation of the
  sachdev-ye-kitaev model by asymmetric qubitization,'' {\em Phys. Rev. A},
  vol.~99, p.~040301, Apr 2019.

\bibitem{https://doi.org/10.48550/arxiv.2110.04778}
T.~Reza, S.~M. Frolov, and D.~Pekker, ``A proposal to extract and enhance
  four-majorana interactions in hybrid nanowires,'' 2021.

\bibitem{Maldacena:2016hyu}
J.~Maldacena and D.~Stanford, ``{Remarks on the Sachdev-Ye-Kitaev model},''
  {\em Phys. Rev. D}, vol.~94, no.~10, p.~106002, 2016.

\bibitem{rosenhaus2019introduction}
V.~Rosenhaus, ``An introduction to the syk model,'' {\em Journal of Physics A:
  Mathematical and Theoretical}, vol.~52, no.~32, p.~323001, 2019.

\bibitem{sarosi2017ads}
G.~S{\'a}rosi, ``Ads $ \_2 $ holography and the syk model,'' {\em arXiv
  preprint arXiv:1711.08482}, 2017.

\bibitem{Chowdhury:2021qpy}
D.~Chowdhury, A.~Georges, O.~Parcollet, and S.~Sachdev, ``{Sachdev-Ye-Kitaev
  Models and Beyond: A Window into Non-Fermi Liquids},'' 9 2021.

\bibitem{Kitaev:2017awl}
A.~Kitaev and S.~J. Suh, ``{The soft mode in the Sachdev-Ye-Kitaev model and
  its gravity dual},'' {\em JHEP}, vol.~05, p.~183, 2018.

\bibitem{Gurau:2017xhf}
R.~Gurau, ``{Quenched equals annealed at leading order in the colored SYK
  model},'' {\em EPL}, vol.~119, no.~3, p.~30003, 2017.

\bibitem{2000PhRvL..85..840G}
A.~{Georges}, O.~{Parcollet}, and S.~{Sachdev}, ``{Mean Field Theory of a
  Quantum Heisenberg Spin Glass},'' {\em Physical Review Letters}, vol.~85,
  pp.~840--843, July 2000.

\bibitem{Cotler:2016fpe}
J.~S. Cotler, G.~Gur-Ari, M.~Hanada, J.~Polchinski, P.~Saad, S.~H. Shenker,
  D.~Stanford, A.~Streicher, and M.~Tezuka, ``{Black Holes and Random
  Matrices},'' {\em JHEP}, vol.~05, p.~118, 2017.
\newblock [Erratum: JHEP 09, 002 (2018)].

\bibitem{Jevicki:2016bwu}
A.~Jevicki, K.~Suzuki, and J.~Yoon, ``{Bi-Local Holography in the SYK Model},''
  {\em JHEP}, vol.~07, p.~007, 2016.

\bibitem{Jevicki:2016ito}
A.~Jevicki and K.~Suzuki, ``{Bi-Local Holography in the SYK Model:
  Perturbations},'' {\em JHEP}, vol.~11, p.~046, 2016.

\bibitem{Tikhanovskaya:2020elb}
M.~Tikhanovskaya, H.~Guo, S.~Sachdev, and G.~Tarnopolsky, ``{Excitation spectra
  of quantum matter without quasiparticles I: Sachdev-Ye-Kitaev models},'' {\em
  Phys. Rev. B}, vol.~103, no.~7, p.~075141, 2021.

\bibitem{Tikhanovskaya:2020zcw}
M.~Tikhanovskaya, H.~Guo, S.~Sachdev, and G.~Tarnopolsky, ``{Excitation spectra
  of quantum matter without quasiparticles II: random $t$-$J$ models},'' {\em
  Phys. Rev. B}, vol.~103, no.~7, p.~075142, 2021.

\bibitem{Gross:2016kjj}
D.~J. Gross and V.~Rosenhaus, ``{A Generalization of Sachdev-Ye-Kitaev},'' {\em
  JHEP}, vol.~02, p.~093, 2017.

\bibitem{Gross:2017hcz}
D.~J. Gross and V.~Rosenhaus, ``{The Bulk Dual of SYK: Cubic Couplings},'' {\em
  JHEP}, vol.~05, p.~092, 2017.

\bibitem{Gross:2017vhb}
D.~J. Gross and V.~Rosenhaus, ``{A line of CFTs: from generalized free fields
  to SYK},'' {\em JHEP}, vol.~07, p.~086, 2017.

\bibitem{Gu:2019jub}
Y.~Gu, A.~Kitaev, S.~Sachdev, and G.~Tarnopolsky, ``{Notes on the complex
  Sachdev-Ye-Kitaev model},'' {\em JHEP}, vol.~02, p.~157, 2020.

\bibitem{Iliesiu:2018fao}
L.~Iliesiu, M.~Kolo\u{g}lu, R.~Mahajan, E.~Perlmutter, and D.~Simmons-Duffin,
  ``{The Conformal Bootstrap at Finite Temperature},'' {\em JHEP}, vol.~10,
  p.~070, 2018.

\bibitem{Maldacena:2016upp}
J.~Maldacena, D.~Stanford, and Z.~Yang, ``{Conformal symmetry and its breaking
  in two dimensional Nearly Anti-de-Sitter space},'' {\em PTEP}, vol.~2016,
  no.~12, p.~12C104, 2016.

\bibitem{doi:10.1063/5.0003145}
X.~Dong, D.~Zgid, E.~Gull, and H.~U.~R. Strand, ``Legendre-spectral dyson
  equation solver with super-exponential convergence,'' {\em The Journal of
  Chemical Physics}, vol.~152, no.~13, p.~134107, 2020.

\bibitem{2002BlumerPhd}
N.~{Blümer}, {\em {Mott-Hubbard Metal-Insulator Transition and Optical
  Conductivity in High Dimensions}}.
\newblock PhD thesis, Universität Augsburg, 2002.

\bibitem{2007PhDT180C}
A.-B. {Comanac}, {\em {Dynamical mean field theory of correlated electron
  systems: New algorithms and applications to local observables}}.
\newblock PhD thesis, Columbia University, New York, Jan. 2007.

\bibitem{PhysRevB.94.195119}
D.~H\"ugel, P.~Werner, L.~Pollet, and H.~U.~R. Strand, ``Bosonic self-energy
  functional theory,'' {\em Phys. Rev. B}, vol.~94, p.~195119, Nov 2016.

\bibitem{PhysRevB.84.075145}
L.~Boehnke, H.~Hafermann, M.~Ferrero, F.~Lechermann, and O.~Parcollet,
  ``Orthogonal polynomial representation of imaginary-time green's functions,''
  {\em Phys. Rev. B}, vol.~84, p.~075145, Aug 2011.

\bibitem{PhysRevB.98.075127}
E.~Gull, S.~Iskakov, I.~Krivenko, A.~A. Rusakov, and D.~Zgid, ``Chebyshev
  polynomial representation of imaginary-time response functions,'' {\em Phys.
  Rev. B}, vol.~98, p.~075127, Aug 2018.

\bibitem{kaye2021discrete}
J.~Kaye, K.~Chen, and O.~Parcollet, ``Discrete lehmann representation of
  imaginary time green's functions,'' 2021.

\bibitem{Sachdev:2015efa}
S.~Sachdev, ``{Bekenstein-Hawking Entropy and Strange Metals},'' {\em Phys.
  Rev. X}, vol.~5, no.~4, p.~041025, 2015.

\bibitem{Klebanov:2016xxf}
I.~R. Klebanov and G.~Tarnopolsky, ``{Uncolored random tensors, melon diagrams,
  and the Sachdev-Ye-Kitaev models},'' {\em Phys. Rev. D}, vol.~95, no.~4,
  p.~046004, 2017.

\bibitem{ArguelloCruzTarnopol}
E.~Arguello~Cruz and G.~Tarnopolsky, ``to be published,''

\bibitem{Kim:2019upg}
J.~Kim, I.~R. Klebanov, G.~Tarnopolsky, and W.~Zhao, ``{Symmetry Breaking in
  Coupled SYK or Tensor Models},'' {\em Phys. Rev. X}, vol.~9, no.~2,
  p.~021043, 2019.

\bibitem{Klebanov:2020kck}
I.~R. Klebanov, A.~Milekhin, G.~Tarnopolsky, and W.~Zhao, ``{Spontaneous
  Breaking of $U(1)$ Symmetry in Coupled Complex SYK Models},'' {\em JHEP},
  vol.~11, p.~162, 2020.

\bibitem{Milekhin:2021sqd}
A.~Milekhin, ``{Coupled Sachdev-Ye-Kitaev models without Schwartzian
  dominance},'' 2 2021.

\bibitem{Milekhin:2021cou}
A.~Milekhin, ``{Non-local reparametrization action in coupled Sachdev-Ye-Kitaev
  models},'' {\em JHEP}, vol.~12, p.~114, 2021.

\bibitem{doi:10.1137/140955835}
N.~Hale and A.~Townsend, ``An algorithm for the convolution of legendre
  series,'' {\em SIAM Journal on Scientific Computing}, vol.~36, no.~3,
  pp.~A1207--A1220, 2014.

\bibitem{JieShenSM}
J.~Shen, T.~Tang, and L.-L. Wang, {\em Spectral Methods. Algorithms, Analysis
  and Applications}.
\newblock Springer Series in Computational Mathematics, Berlin: Springer, 2011.

\end{thebibliography}

\end{document}